\documentclass[10pt,conference]{IEEEtran}

\usepackage[utf8]{inputenc}
\usepackage[T1]{fontenc}
\usepackage{graphicx}
\usepackage{ifthen}
\ifCLASSOPTIONcompsoc
  \usepackage[caption=false,font=normalsize,labelfont=sf,textfont=sf]{subfig}
\else
  \usepackage[caption=false,font=footnotesize]{subfig}
\fi
\usepackage{caption}
\usepackage[abbreviations]{foreign}
\usepackage{pifont}
\usepackage{flushend}
\usepackage{fontawesome}
\usepackage[binary-units,per-mode=symbol]{siunitx}
\usepackage{tikz}
\usepackage{xspace}
\usepackage[sort]{natbib}
\usepackage[colorlinks=false,linkcolor=blue,urlcolor=blue,citecolor=blue,bookmarks=false,draft]{hyperref}
\graphicspath{{./}{figures/}}
\setcitestyle{square,numbers}

\newcommand\copyrighttext{  \footnotesize \textcopyright 2018 IEEE.
    Personal use of this material is permitted.
    Permission from IEEE must be obtained for all other uses,
    in any current or future media, including reprinting/republishing this
    material for advertising or promotional purposes, creating new collective
    works, for resale or redistribution to servers or
    lists, or reuse of any copyrighted component of this work in other works.
    Presented in the
    \href{http://www.lasid.ufba.br/srds2018/view/index.php}{37th IEEE
      International Symposium on Reliable Distributed Systems (SRDS
      '18)}. The final version of this paper is available under DOI: \href{https://doi.org/10.1109/SRDS.2018.00042}{10.1109/SRDS.2018.00042}}
\newcommand\copyrightnotice{\begin{tikzpicture}[remember picture,overlay]
\node[anchor=south,yshift=10pt,fill=yellow!20] at (current page.south) {\fbox{\parbox{\dimexpr\textwidth-\fboxsep-\fboxrule\relax}{\copyrighttext}}};
\end{tikzpicture}}

\begin{document}

\title{Security, Performance and Energy Implications of
  Hardware-assisted Memory Protection Mechanisms on Event-based Streaming Systems}
\date{}
\author{
\IEEEauthorblockN{Christian~Göttel,
  Rafael~Pires,
  Isabelly~Rocha,
  Sébastien~Vaucher,
  Pascal~Felber,
  Marcelo~Pasin,
  Valerio~Schiavoni}
\IEEEauthorblockA{University of Neuch{\^a}tel, Switzerland ---
  \texttt{first.last@unine.ch}}}

\maketitle
\copyrightnotice

\section{Hardware-assisted Memory Protection Mechanisms}
Major cloud providers such as Amazon~\cite{amazonskylake},
Google~\cite{gceskylake} and Microsoft~\cite{azureconfidential} 
provide nowadays some form of infrastructure as a service (IaaS) which allows
deploying services in the form of virtual machines~\cite{amazon-ec2}, containers~\cite{google-kub} or
bare-metal~\cite{amazon-baremetal} instances. Although software-based
solutions like homomorphic encryption exit, privacy
concerns~\cite{pearson2010privacy} greatly hinder the deployment of such 
services over public clouds. It is particularly difficult for
homomorphic encryption to match performance requirements of modern
workloads~\cite{naehrig2011can}.
Evaluating simple operations on basic data types with HElib~\cite{halevi2013design}, a homomorphic
encryption library, against their unencrypted counter part reveals,
that homomorphic encryption is still impractical under realistic workloads.

In recent attempts to enable privacy-preserving operations,
publish/subscribe systems among other types of communication services
have received much attention. Meanwhile, Intel and AMD have introduced
hardware-assisted memory protection mechanisms inside x86 processors
to provide answers overcoming the limitation of current software-based
solutions.
With the launch of the Skylake generation, Intel added a new technology
called Software Guard Extension (SGX)~\cite{costan_intel} to their 
processors. SGX allows applications to create secure \emph{enclaves}
protecting the confidentiality and integrity of data and its
associated code during execution.
An application has to be signed and shipped as an unencrypted shared
library (respectively a shared object on Linux systems) in order to be executed
in an enclave. Execution of an enclave on a genuine Intel processor
with enabled SGX technology can be ensured by a \emph{remote attestation
  protocol}. The enclave is stored in the \emph{enclave page
  cache} (EPC) when executed; a limited memory area predefined at boot
time. Page eviction is handled by the SGX driver and confidentiality,
integrity, replay and tamper protected by the \emph{memory encryption
  engine} (MEE)~\cite{gueron2016memory}.
AMD's recently introduced Zen microarchitecture is capable of
transparently encrypting memory pages using their novel technologies Secure
Memory Encryption (SME) and Secure Encrypted Virtualization (SEV). SME
and SEV make use of \emph{ephemeral} encryption keys required by an AES engine
located on the core's memory controller. While SME creates a single key
to encrypt the entire system's memory, SEV can generate and assign a key
to a limited number of distinct virtual machines and a single hypervisor
running on the processor. The creation of the memory
keys is delegated to the \emph{secure processor} (SP), an ARM
\textsc{TrustZone}-enabled system-on-chip (SoC) embedded on-die~\cite{kaplan2016amd}.

\begin{figure}[t!]
	\centering  
	\includegraphics[scale=0.8]{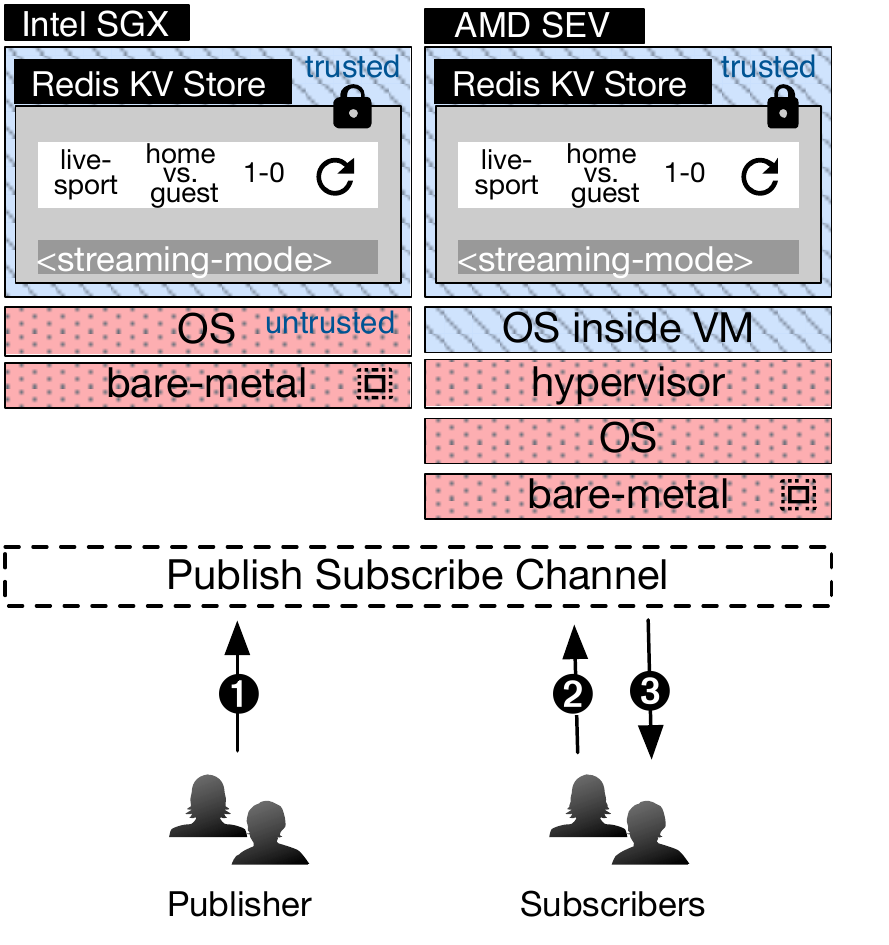}
	\captionsetup{belowskip=-10pt}
	\caption{Architecture of our system and differences when deployed with Intel SGX and AMD SEV-ES. The components with a diagonally hatched pattern on a blue background are trusted, those with a dotted red background are untrusted, respectively. Redis is configured in \emph{streaming-mode}~\cite{redisstreaming}.\label{fig:architecture}}
\end{figure}

\section{Architecture}
\label{sec:architecture}

We designed and implemented a simple yet pragmatic event-based streaming
system to evaluate our execution. The core of our system consists of a
key-value store with native support to register and trigger callback
functions associated to CRUD operations (\ie create, read, update,
delete) on key-value entries. These callback functions implement
matching filters for subscribers of a publish/subscribe system, that
will receive events upon notification of the channel.

\begin{figure*}[!t]
	\centering
	\subfloat{%
		\includegraphics[width=\columnwidth]{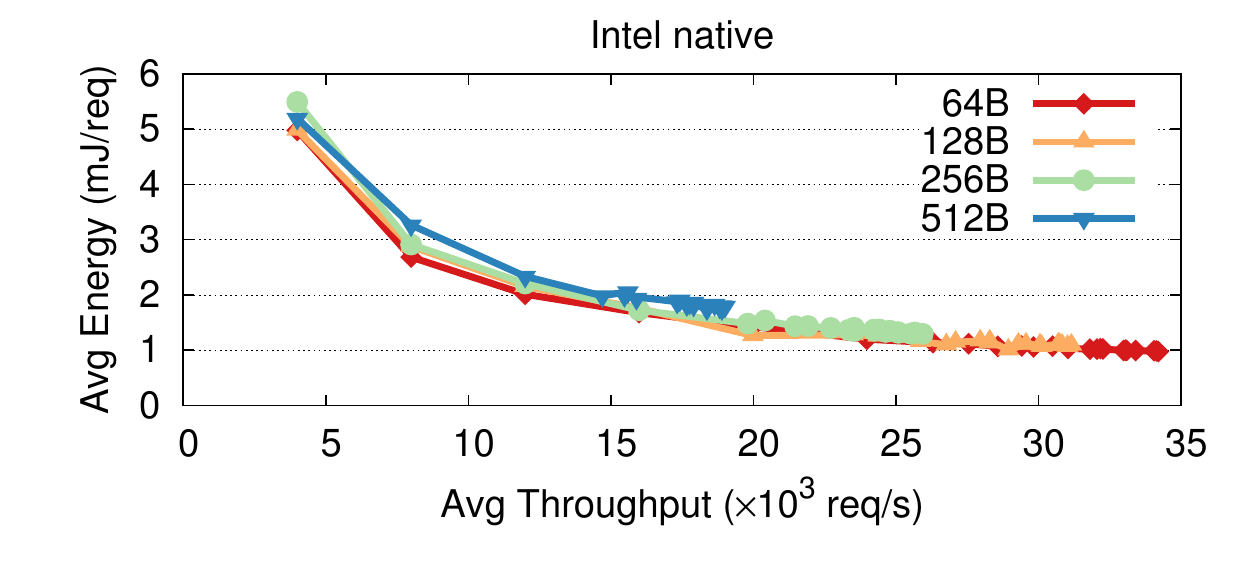}
		\label{fig:pubsub:energy:intel}
	}%
	\subfloat{%
		\includegraphics[width=\columnwidth]{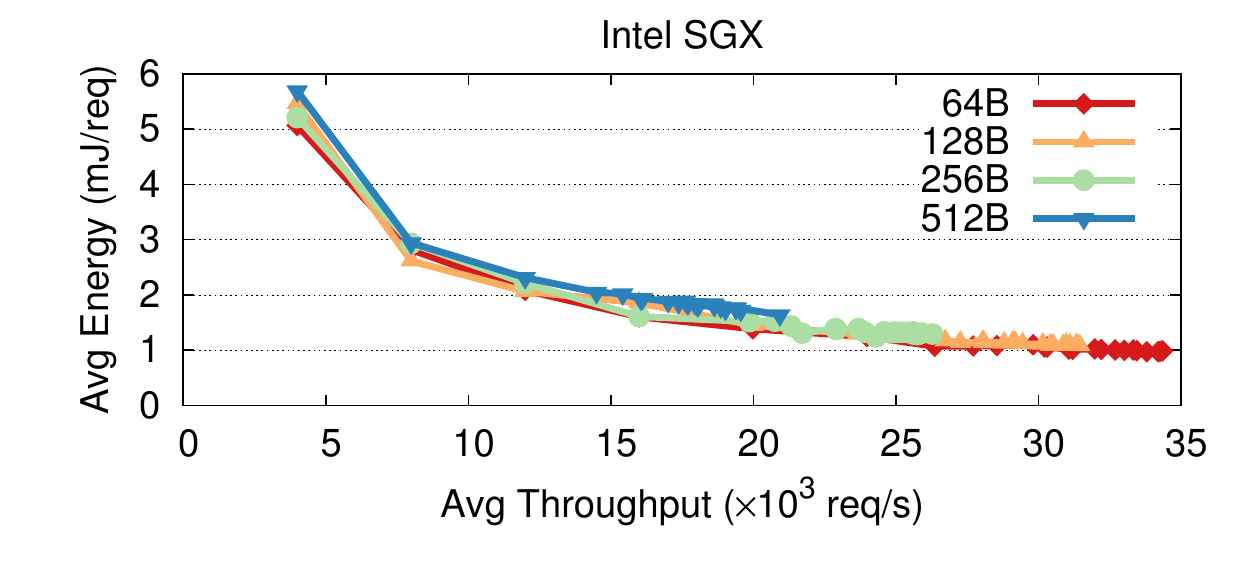}
		\label{fig:pubsub:energy:sgx}
	}
	\\[-12pt]
	\subfloat{%
		\includegraphics[width=\columnwidth]{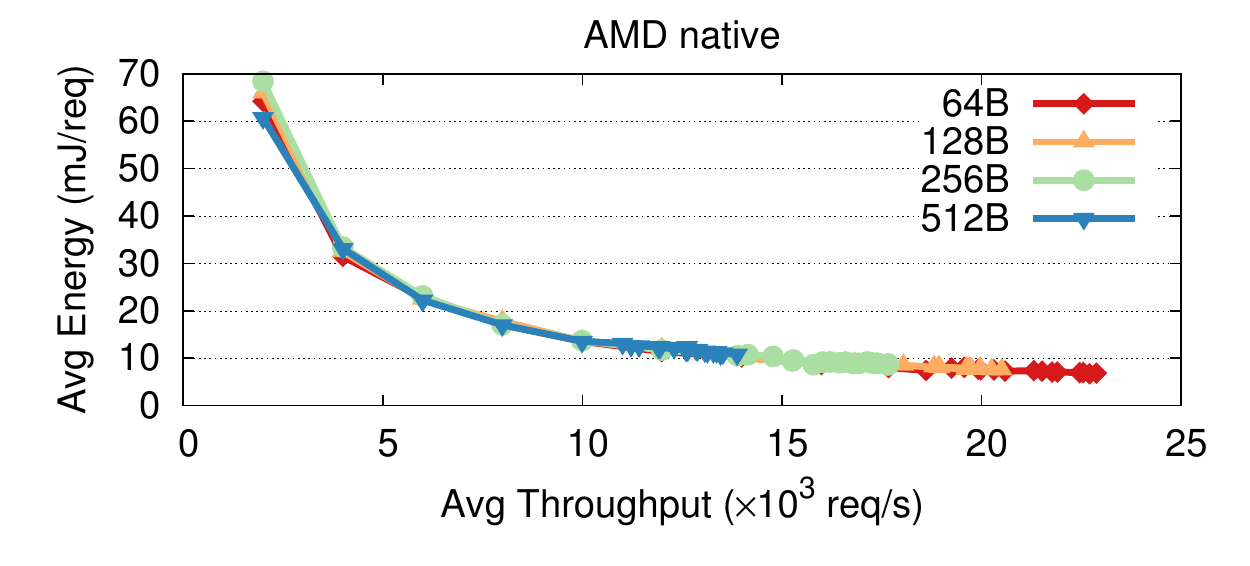}
		\label{fig:pubsub:energy:amd}
	}%
	\subfloat{%
		\includegraphics[width=\columnwidth]{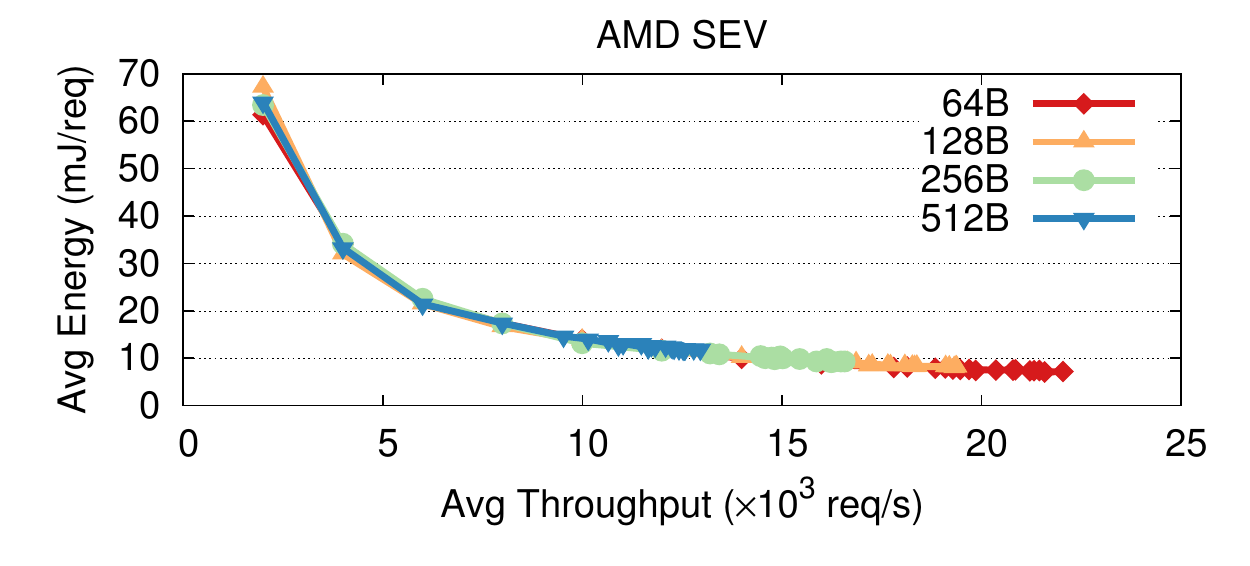}
		\label{fig:pubsub:energy:sev}
	}
	\captionsetup{skip=-3pt,belowskip=-12pt}
	\caption{Macro-benchmark: energy cost of publish/subscribe.}
    \label{fig:pubsub:energy}
\end{figure*}

The main components of the event-based streaming system are depicted
in~\autoref{fig:architecture} with Intel SGX (left) and AMD SEV
(right). All potentially sensitive components \ie the key-value store
and its content, the callback functions and the endpoints of the
publish/subscribe channels, have to be protected by SGX or SEV. However,
only the key-value store entries are considered to be protected by SGX or
SEV in our implementation. Furthermore, we do not include other
additional stages in the processing pipeline nor do we explicitly
include broker or broker overlays~\cite{eugster2003many}. Under these
carefully controlled conditions side-effects of SGX and SEV can be
better highlighted on the main processing node, in particular
memory-bound operations and their energy cost.

The workflow of operations is as follows.
First, a subscriber manifests its interests by subscribing to the channel (\autoref{fig:architecture}-\ding{202}).
Then, publishers start emitting events with a given content, \eg, the results of a sport event (\autoref{fig:architecture}-\ding{203}).
As soon as the content is updated, a callback function is triggered (\autoref{fig:architecture}-\faRepeat).
Finally, the potential subscriber(s) receive the event (\autoref{fig:architecture}-\ding{204}).

\section{Implementation}
The architecture was implemented on top of well-known open-source
systems and libraries. Redis~\cite{redis} (v4.0.8), an efficient and
lightweight in-memory key-value store, is used as core
component. It also features a built-in publish/subscribe
system, which is exploited in order to realize our experimental
platform. Publishers and subscribers connect to the channels provided by
Redis using Jedis~\cite{jedis} (v2.9.0) Java bindings for Redis.
The callback system is implemented by leveraging Redis' ability to load
external modules~\cite{redis:modules}.
Despite Redis being a single threaded application, modules can be
run in a multi-threaded setup.

While applications run under AMD SEV do not require any changes, they do
need to be modified under Intel
SGX. Graphene-SGX~\cite{tsai2017graphene} is a library allowing to run
unmodified applications inside enclaves and was used for this
benchmark. In order to run unmodified applications under SGX, a manifest
file has to be provided to Graphene-SGX specifying the resources, \eg
shared libraries, files, network endpoints, which the enclave is allowed
to make use of. The manifest file is pre-processed by an auxiliary tool
generating signatures that are later checked by the Graphene loader.

Various workloads were injected using
YCSB~\cite{cooper2010benchmarking}, v0.12.0 commit \texttt{3d6ed690}, to
record latency, throughput, performance and energy values.

\section{Evaluation}

The benchmark measures the latency from the moment a publisher emits a
new event until the moment all subscribers receive the content of the
event.
Four different configurations of the system are evaluated:
\emph{(i)}~Intel without SGX protection,
\emph{(ii)}~with SGX by leveraging Graphene,
\emph{(iii)}~AMD without memory protection and
\emph{(iv)}~AMD with SEV.
Publishers are configured such that they inject new events at
fixed throughputs with fixed message sizes ranging from \SI{64}{\byte}
up to \SI{512}{\byte}.

Measured latencies for smaller message sizes are consistently lower for
higher throughputs (requests/second).
The cost of serialization reduces the efficiency of our
system for larger message sizes.
A pairwise comparison of the configurations for Intel and
AMD reveals how these memory protection mechanisms are negatively
affecting the observed latencies, which is particularly evident for
Intel configurations. This observation is further confirmed by bandwidth
usage values.

The energy cost of messages send over the publish/subscribe system
is shown in~\autoref{fig:pubsub:energy}. As the system begins to occupy a
significant amount of the machine's resources, energy consumption begins
to increase at a linear rate relative to the target throughput. This
behavior is reflected by the decreasing energy cost per message before
reaching a minimal energy cost.

Memory requirements do not exceed the available EPC for the Intel
configurations under these evaluation settings. Consequently, the
measurements indicate that both memory protection mechanisms consume an
even amount of energy compared to their native setup.

Due to the energy being recorded for the entire system using an external
power measurement device, it becomes difficult to make an
implication on the energy consumption of the memory protection mechanisms.
In a next step we are developing tools to measure the energy consumption
of precesses at a much finer grained level, for instance  the processor's
core. Such tools would then give us the opportunity to observe in more
detail the influence of memory protection mechanisms on processes and
assist in the development of novel security and energy-aware system components.

\section*{Acknowledgments}
The authors would like to thank Christof Fetzer for the discussions on hardware-assisted memory protection mechanisms.
The research leading to these results has received funding from the
European Union's Horizon 2020 research and innovation programme under
the LEGaTO Project
(\href{https://legato-project.eu/}{legato-project.eu}), grant agreement
No~780681.

\bibliographystyle{IEEEtran}
\bibliography{IEEEabrv,biblio}

\end{document}